\documentclass[aps,pre,preprint,epsfig,showpacs]{revtex4}
\usepackage{epsfig}
\usepackage{graphicx}
\begin{document}

\title{Filament attachment dynamics in actin-based propulsion}

\author{J. I. Katz}
\email{katz@wuphys.wustl.edu}
\thanks{Work supported by NIH grant GM38542}
\affiliation{Department of Physics and McDonnell Center for the Space Sciences,
Washington University, St.~Louis, Mo. 63130}

\author{A. E. Carlsson}
\email{aec@wuphys.wustl.edu}
\thanks{Work supported by NIH grant GM38542}
\affiliation{Department of Physics, Washington University, St.~Louis, Mo. 63130}

\date{\today}

\begin{abstract}
Theory and experiment have established that F-actin filaments are strongly
attached to the intracellular parasites (such as {\it Listeria}) they propel
with ``comet tails''.  We consider the implications of these observations for
propulsion.  By calculating the motion produced in various models of
attachment and comparing to experiment we demonstrate that the attachment must
be sliding rather than hinged.  By modeling experiments on ActA-coated spheres
we draw conclusions regarding the interaction between F-actin and their
surfaces that may also be applicable to living systems.
\end{abstract}

\pacs{87.15.Kg, 87.16.Ac, 87.17.Aa, 87.17.Jj}

\maketitle

\section{Introduction}

A number of intracellular parasites, some of medical as well as scientific
interest, propel themselves through host cells by suborning the host's actin
polymerization machinery, inducing it to provide propulsive force for the
parasite~\cite{C95,T95,DC98,G99,H99,B00,BS00,C00,C01,T00,PLC01}.  These
parasites include {\it Listeria monocytogenes, Shigella flexneri,
Rickettsiae} and {\it Vaccinia} virus.  Catalytic proteins on the surface of
the parasite initiate the growth of new polymeric (F-) actin filaments on the
surface.

F-actin appears to be tightly bound to the surface it pushes.  Evidence for
binding~\cite{G00a,G00b,KM00,OH00,RG01,DP02,MO03} includes measurement of
discrete displacement steps nearly equal to the diameter (5.4 nm) of G-actin,
direct observation and the theoretical argument that in the absence of binding
Brownian diffusion would readily sever the contact between the parasite and
its propelling bundle of F-actin filaments (called a comet-tail, from its
micrographic appearance).  Angular diffusion is also rapid in the absence of
binding; for a sphere of radius $r = 1\,\mu$m, $D_\theta = kT/8\pi \eta r^3 =
0.16$ radian$^2$/sec, taking the viscosity $\eta$ of pure water; the effective
viscosity of cytoplasm is greater, scale dependent and poorly
understood~\cite{M98,B99,LP00}.   Even if the effective viscosity were 100
times greater, unattached bacteria would tumble rapidly, in contradiction to
observation.

Maintaining contact over a time $t$ requires an attractive potential $V \le
- kT \ln{(\nu t)}$, where $\nu$ is a relaxation rate (an effective attempt
frequency).  A number of processes contribute to $\nu$: viscous drag on the
sphere, the (damped) elastic modes of the actin filaments, and elastic/plastic
flow of the actin gel embedded in the surrounding aqueous medium.  Of these,
only the first is known quantitatively: The viscous damping of a $\mu$m-sized
object of density $\rho$ and radius $r$ gives $\nu = 9 \eta/(2 \rho r^2) \sim
10^7$ sec$^{-1}$ (in water).  Taking $t \ge 10^3$ sec as an empirical lower
bound on the attachment time we find $V \le - 1.0 \times 10^{-19}$
J~\cite{OH00}.  Because the dependence of $V$ on $\nu$ is logarithmic, this
result is only weakly dependent on uncertainties (such as the applicable
viscosity) in $\nu$.

The interaction between a filament of F-actin and the protein-covered
surface to which it is bound is complex, and not calculable from {\it ab
initio} interatomic potentials.  The purpose of this work is to constrain
that interaction by calculating the consequences for propulsion of simple
models of the interaction, and then comparing the results to experiment.

Intercalation is driven by the free energy~\cite{GYK76} $\Delta G \approx 6
\times 10^{-20}$ J released when a molecule of G-actin is added to a filament
of F-actin.  During intercalation a single filament exerts a force $F \sim
\Delta G/a \approx 20$ pN on the G-actin, drawing it into the gap between the
F-actin and the surface proteins to which it is bound.  This force is
sufficient to drive $\mu$m-sized objects at speeds $\sim 0.1$ cm/sec against
viscous drag (taking the viscosity to be that of water), so the intercalation
is complete and $\Delta G$ dissipated in a few $\mu$sec.  The product of the
duration of a single intercalation $t_I = 6 \pi \eta a^2 r/\Delta G$ (the
force must move the sphere and the filament to make room for the intercalated
monomer) and the Brownian relaxation rate $\nu$ of a propelled sphere defines
a new dimensionless number which we call the intercalation smoothness
\begin{equation}
N_{IS} \equiv t_I \nu = {27 \pi \eta^2 a^2 \over \Delta G \rho r}.
\end{equation}
The intercalating Reynolds number of both the sphere and the G-actin are
\begin{equation}
{\rm Re} = {9 \over 2} {a \over r} {1 \over N_{IS}} \ll 1.
\end{equation}
When $N_{IS} \gg 1$ and ${\rm Re} \ll 1$, as is the case here ($N_{IS}
\approx 10^2$ for $r = 0.25\,\mu$m, taking the viscosity of pure water) the
Stokes flow approximation (implicitly averaging over the sphere's and
filament's Brownian motion) may be used during the intercalation, even though
the entire intercalation is effectively instantaneous compared to the
intervals between intercalations.  $N_{IS}$ is related to the Reynolds number,
but not entirely determined by it because of the additional factor $a/r \sim
10^{-2}$.  There are parameter regimes (not relevant to the experiments
discussed here) for which both ${\rm Re} \ll 1$ and $N_{IS} \ll 1$, so that
Stokes flow would be applicable but in which it would not be valid to average
over the Brownian motion of the sphere or filament.

\section{Calculations}

We have performed Monte-Carlo simulations of the effects of G-actin
intercalation in a model in which the F-actin is attached to the surface of a
sphere.  This model is necessarily much simplified compared to a full physical
description.  For example, we employ crude approximations to the flow around
the sphere with attached F-actin and ignore cross-linking within the actin
tail.  Recognizing the crudity of our models, we note that doing better would
require either formidable numerical calculations (for example, of the flow
around a sphere with attached filaments) and quantitative understanding (of
actin cross-linking, and of the precise geometry of the attached filaments)
which does not exist.  Despite these rough approximations, we believe our
qualitative conclusions are reliable and useful. 

We use three-dimensional physics except for a model of the geometry in which
the F-actin is constrained to lie in a single equatorial disc.  We take the
flow fields to be those of Stokesian flow around a sphere.  The surrounding
fluid, as is generally the case in low Reynolds number
hydrodynamics~\cite{HB83}, is an effectively infinite sink of momentum and
angular momentum, just as it is also a heat bath.  

The crucial question is the mechanism by which the
symmetry of a particle initially uniformly covered with intercalation sites
is broken, producing directed motion.  Symmetry is much simpler to define
and easier to achieve on the rim of a disc (on which equally spaced points
are equivalent) than on the surface of a sphere (on which it is not, in
general, possible to distribute $N$ equivalent points).  Symmetry-breaking
results from amplification of statistical fluctuations in the locations of
the intercalation sites, which depend on $N$ but not on dimensionality.

At the beginning of a Monte Carlo run $N$ ($N = 50$ in the calculations
shown here), intercalation sites are distributed uniformly around the rim of
the disc.  By eliminating any statistical deviation from symmetry in the
location of sites we focus attention of the mechanism by which symmetry is
broken.  The loci of intercalation events are chosen randomly from these
sites.  The time scale is arbitrarily defined by an assumed intercalation
rate, but all other parameters are physically meaningful.  Each intercalation
introduces a relative displacement of $a = 2.7$ nm between the F-actin
(initially containing zero monomers) and its attachment point.  This
displacement is divided between the filament and the sphere in inverse
proportion to their viscous drags (using the three dimensional results for an
isolated sphere without attached filaments and a prolate
ellipsoid~\cite{B83,HB83}).  Their mutual hydrodynamic interaction and their
interactions with the other filaments are small and not calculable
analytically, and are ignored.  The displacement of the attachment point is
resolved into a radial part, which displaces the disc, and a tangential part,
which rotates it.

When the disc is displaced it is surrounded by a Stokes flow field (taken to
be that of a sphere).  All filaments are immersed in this flow field, affect
it, and move with it.  A quantitative calculation of the flow field would not
be feasible in this complex geometry, so we approximate it by assuming each
filament, if free to rotate, is rotated about its attachment point by an
angle $\Delta \theta = fa \sin\phi/(r + \ell/2)$, where $f$ is the fraction of
the relative displacement accommodated by the disc, $\ell$ is the length of the
filament, $r = 0.25\,\mu$m is the sphere's radius and $\phi$ is the angle
between the radius vector to the attachment point and the sphere's displacement
vector.  This approximates moving the midpoint of the (nearly rigid) filament
along with its local (Stokesian) flow field around a sphere.  If the
attachment points are permitted to slide along the sphere's periphery they are
displaced by an angle (measured at the center of the equatorial disc) $\Delta
\theta = fa \sin\phi / (r + \ell/2)$; a small minimum angular separation
between filaments is imposed.  We consider only a single particle in an
infinite fluid; this amounts to requiring that any walls or other particles or
filaments are many times more distant than the diameter of the sphere or the
length of the filament.

Similarly, when the sphere is rotated there is a surrounding
Stokesian~\cite{HB83} flow field.  We approximate the rotation of each
filament about its attachment point by an angle $\Delta \theta = (3/4)fa
\sin\theta_t/(r + \ell/2)$, where the numerator is the displacement of the
disc circumference and $\theta_t$ is the initial angle between the filament
and the normal at its attachment point.

We first consider a model in which the attachments are fixed hinges.
Initially, intercalation produces only infinitesimal displacement of the
sphere because as $\ell \to 0$, $f \to 0$.  As the filaments lengthen $f \to
1$ and the sphere's displacement at each intercalation increases.  The
rotation of the filaments around their hinges, resulting from their viscous
drag, now leads to their being swept back in a direction opposite to the
sphere's motion.  Once swept back, further intercalation tends, on average, to
propel the sphere in the direction of its earlier motion.  The symmetry of the
initial conditions is broken by the stochastic fluctuations in the
intercalation sites, and the sphere acquires a systematic motion.  Increasing
orientational asymmetry of the filaments leads to preferential motion in the
direction opposite to the mean filament, and further orientation of the
filaments in that direction as the process runs away.  The Reynolds number
remains small, so this is not an inertial effect; the directional memory
resides in the orientation of the filaments.

In a second model the attachment points are free to slide along the surface
of the sphere as they are swept back by the Stokes flow, but are held
perpendicular to the sphere, as is considered, for example, by~\cite{DP02}
(and is achieved in simple mechanical devices).  No torques are exerted on the
sphere so it does not systematically rotate.  Once statistical fluctuations
break the initial symmetry, intercalation between swept-back filaments and the
sphere preferentially pushes the sphere in the direction of its earlier
motion.  This process then runs away.  As before, the Reynolds number remains
very small and the directional memory resides in the locations of the
filaments' attachment points.

\section{Results}

\begin{figure}
\centerline{\epsfig{figure=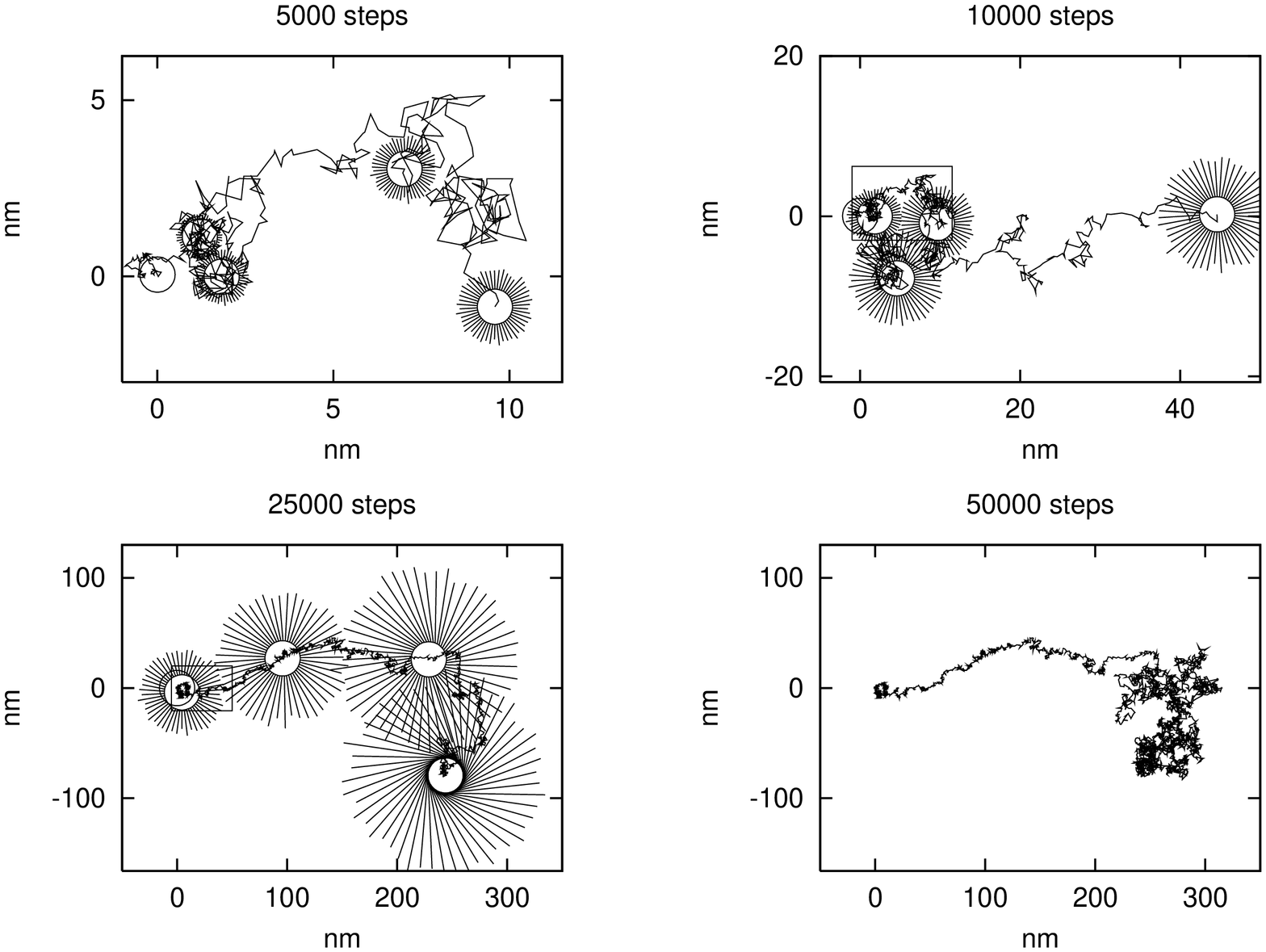,angle=270,width=6.5in}}
\caption{\label{fig1}The motion of a representative sphere with hinged
attachments.  The ``sunbursts'' show the sphere and attached filaments (the
relative dimensions of sphere and filaments are correct, but not on the same
scale as the paths).  The rectangles in some frames show the sizes of the
previous frames for comparison.  An initial random walk soon yields to
directed motion, but around step 20000 runaway ``pinwheel'' rotation becomes
dominant.  The rotation at the hinges is evident in the sunburst for step
25000.  The last frame shows the end of directed motion.}
\end{figure}

\begin{figure}
\centerline{\epsfig{figure=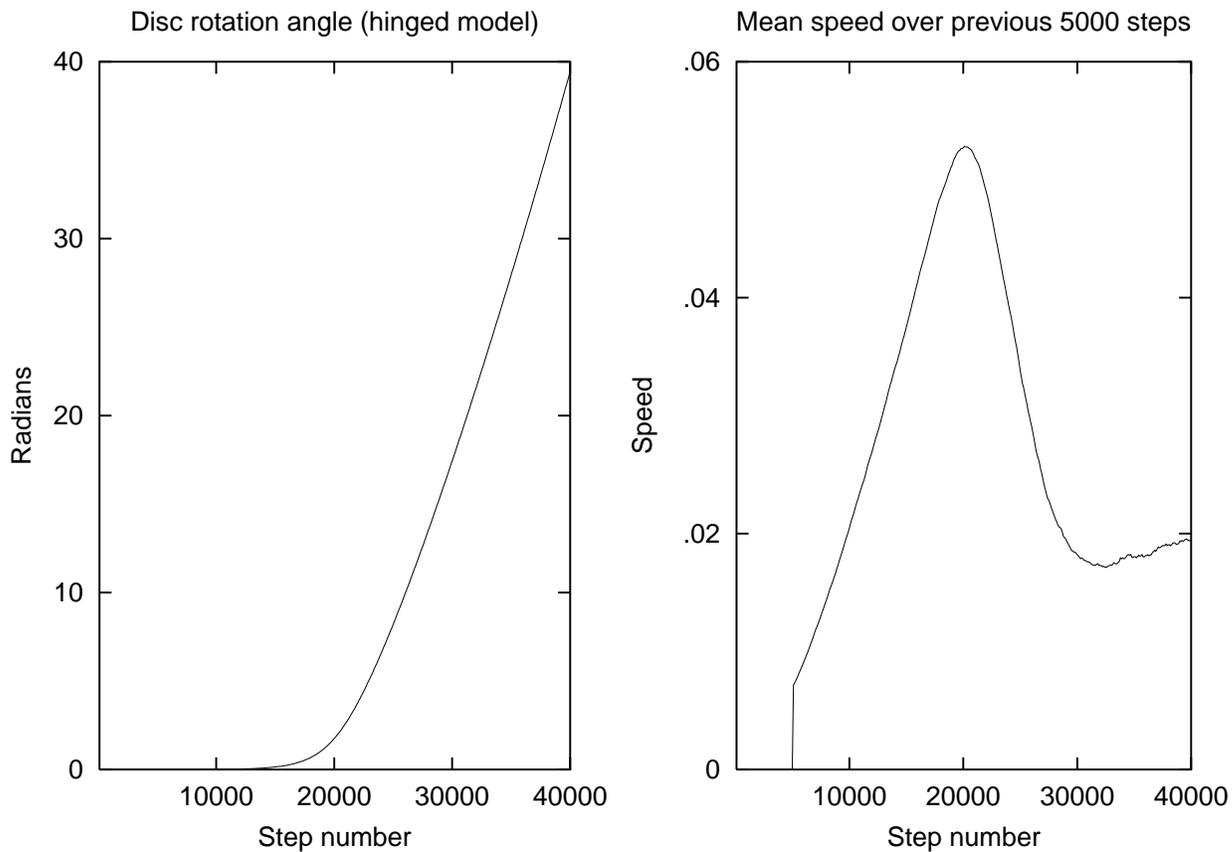,angle=270,width=6.5in}}
\caption{\label{fig2}Averages (over 1000 runs) of the absolute value of the
disc rotation angle and mean speed (arbitrary units) for hinged attachments.
Runaway ``pinwheel'' rotation beginning around step 20000 is evident, with a
corresponding decrease in mean speed.  The mean filament length is the step
number divided by the number of filaments, 50 in this calculation.}
\end{figure}

Results of numerical simulation of the hinged model are shown in
Fig.~\ref{fig1} and Fig.~\ref{fig2}.  Initially random fluctuations first
give way to directed displacement.  Later, rotation runs away as it sweeps
the filaments back around their hinges, and this orientation contributes to
further rotation like a pinwheel.  The mean speed then drops, directed
displacement ends, and no comet-tail forms.

\begin{figure}
\centerline{\epsfig{figure=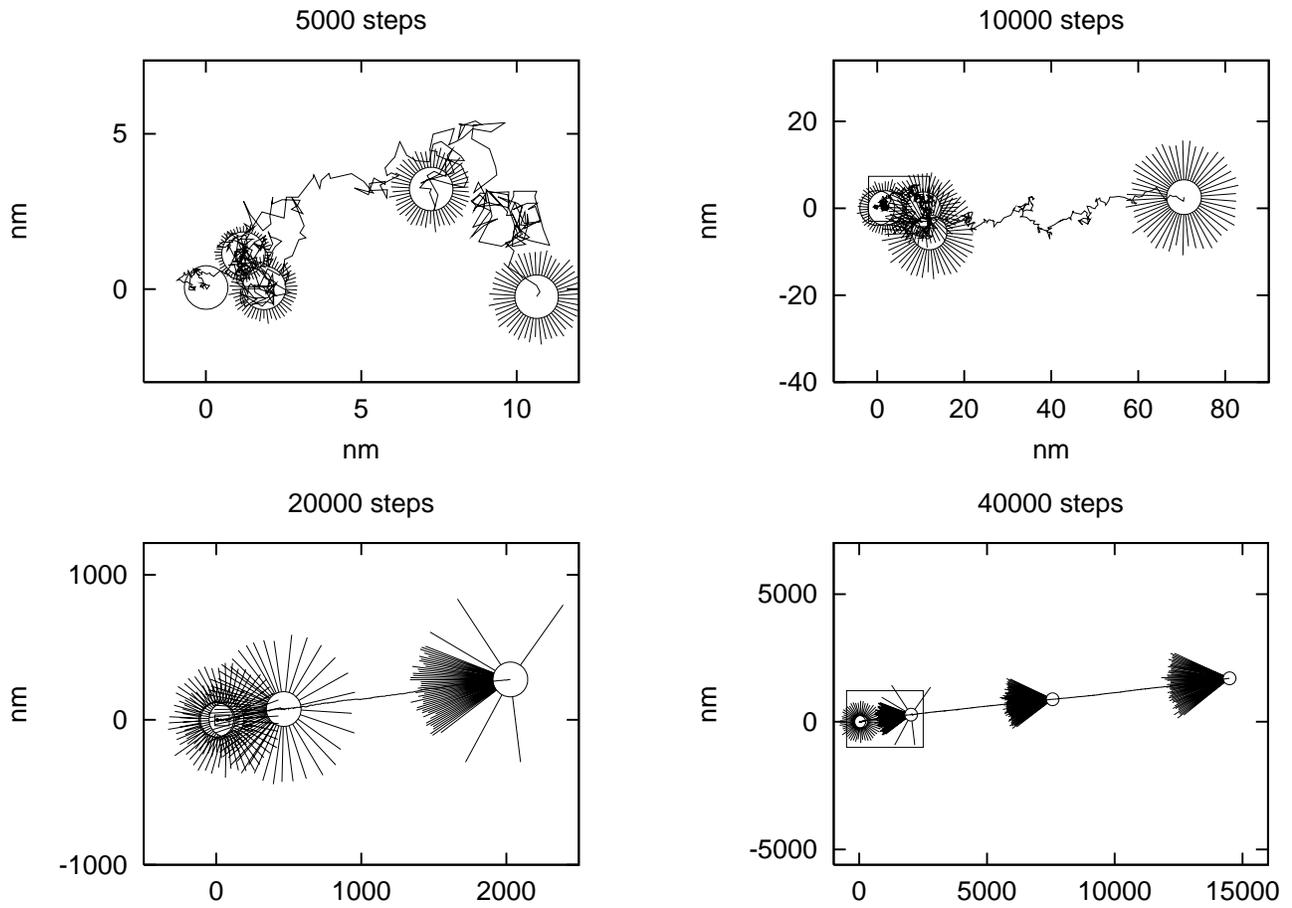,angle=270,width=6.5in}}
\caption{\label{fig3}The motion of a representative disc with sliding normal
attachments.  ``Sunbursts'' and rectangles are as in Fig. 1.  Initially
random motion gives way to directed motion and the formation of a comet-tail.
An arbitrary minimum angle between filaments prevents the collapse of the
comet tail to a single line.}
\end{figure}

\begin{figure}
\centerline{\epsfig{figure=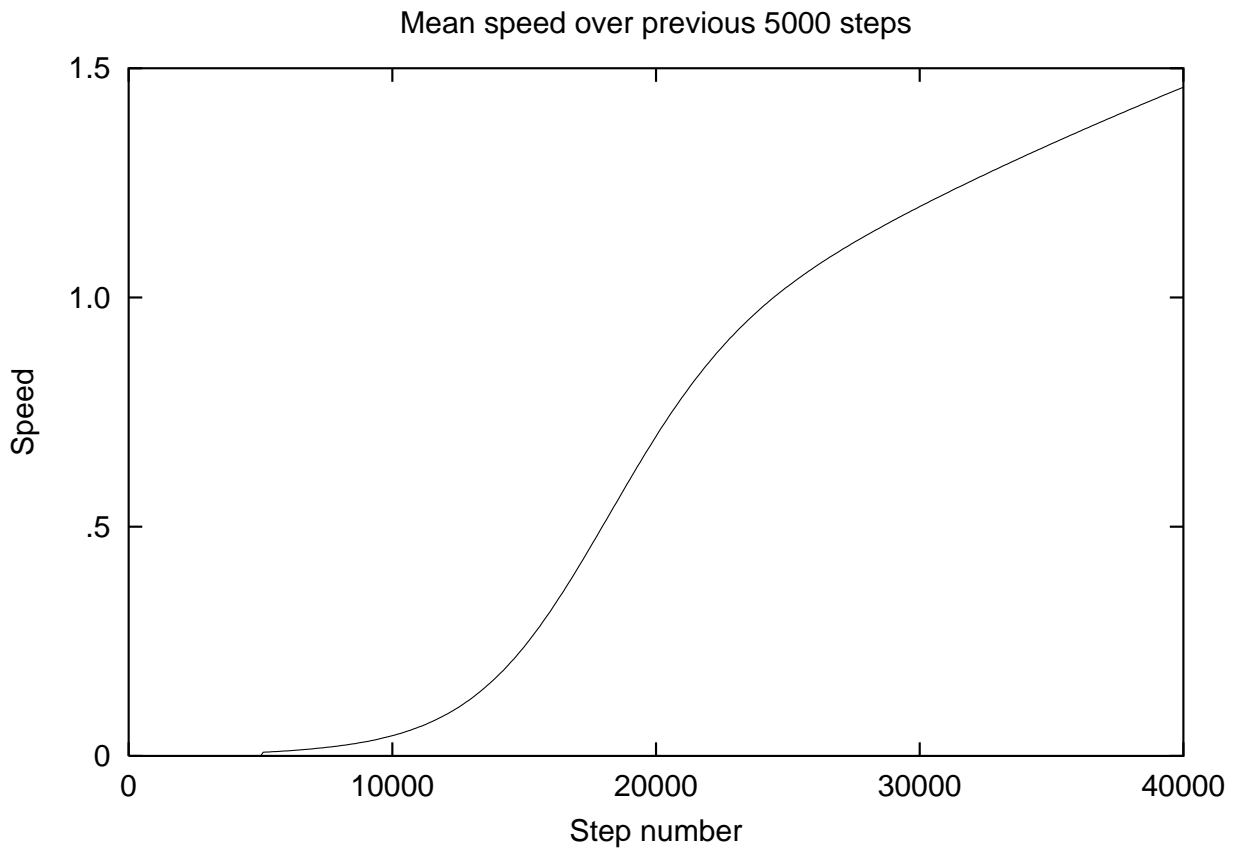,angle=270,width=6.5in}}
\caption{\label{fig4}Average (over 1000 runs) of the mean disc speed with
sliding attachments, showing acceleration as the filaments are swept back into
a coherent comet-tail.  The speed is in the same units as, and should be
compared to, Fig. 2; it continues to increase $\propto f \to 1$ after a comet
tail is formed around step 20000.}
\end{figure}

Results for the sliding model are shown in Fig.~\ref{fig3} and
Fig.~\ref{fig4}.  The random number generator was initialized with the same
seed as in Fig.~\ref{fig1} and Fig.~\ref{fig2}; the initial motion is very
similar because little rotation or translation of the filaments has occurred
in either model.  However, between 7500 and 10000 steps (the fourth and
fifth ``sunbursts'' in the upper right panel of Fig.~\ref{fig3}) the
attachments have slid significantly and directed motion has begun.  Soon
thereafter, this process runs away, the filaments condense to an ordered
comet-tail, and the motion becomes steady and directed.  Within the limits
of the computational model (which has no surrounding cytoskeleton or
branching or cross-linking of filaments), this is a satisfactory
representation of the observed comet-tails.

\section{Discussion}

These results may help explain the experiments of~\cite{C99,OT99,C04}, in which
polystyrene spheres partially but spherically symmetrically (as well as can
be achieved experimentally) coated with the actin polymerization-stimulating
protein AcTa and immersed in cytoplasm ({\it Xenopus} egg extract) were
observed after a latency time to break their initial symmetry and develop
comet-tails of F-actin and directed motion.  This behavior is similar to
that which we find with sliding attachments (as pointed out in~\cite{RG01},
in this experiment the ActA may ``crawl'' on the surface of the beads).

The most remarkable feature of these experiments is the
non-monotonic dependence of bead motility (and comet-tail formation) on the
fraction of bead surface covered by ActA, peaking around 3/8 coverage.  This
is naturally explained by our calculations, for a bead sparsely covered with
ActA will develop little propulsive force (bead motion is restrained by
pre-existing cytoskeleton), while a bead completely covered offers no room
for the ActA, and attached F-actin, to be swept back into a comet-tail.
Thus, from these experiments and our calculations we determine the
properties of the F-actin binding to the bead surface and constrain
microscopic mechanical models such as those of~\cite{DP02}.  In contrast,
experiments~\cite{N00} in which ActA is covalently bound to beads do not
show comet tails and propulsion, which is attributable to the inability of
covalently bound ActA to slide over the beads' surfaces.

Latency was also found in experiments~\cite{BG02} on spheres in a synthetic
growth medium.  In these experiments beads continuously covered with
actin did move, but in a saltatory manner, apparently as a result of elastic
stresses~\cite{S04} (not considered here) in a fractured continuous shell of
F-actin.  When the coverage was only partial, the motion was continuous,
resembling the results of~\cite{C99} and agreeing with our calculations.

\begin{acknowledgments}
We thank S. Block, J. Cooper, A. Mogilner, D. Sept and J. Theriot for useful
discussions.
\end{acknowledgments}

\end{document}